\documentclass[aps,prl,a4paper,twocolumn,showpacs,superscriptaddress,floatfix]{revtex4}
\usepackage[T1]{fontenc}
\usepackage[english]{babel}
\usepackage{dcolumn}
\usepackage{bm}
\usepackage{rotating}
\usepackage{lscape}
\usepackage{stmaryrd}
\usepackage{latexsym}
\usepackage{amsmath}
\usepackage{graphics}
\usepackage{graphicx}
\usepackage{bm}
\usepackage{amssymb}
\usepackage{color}
\usepackage{fancybox}
\usepackage{amsfonts}
\usepackage{afterpage}
\usepackage{epsfig}
\allowdisplaybreaks
\definecolor{MyDarkBlue}{rgb}{0,0.08,0.45}

\newcommand{\su}{\uparrow}
\newcommand{\giu}{\downarrow}

\newcommand{\be}{\begin{equation}}
\newcommand{\ee}{\end{equation}}
\newcommand{\bea}{\begin{eqnarray}}
\newcommand{\eea}{\end{eqnarray}}
\newcommand{\ba}{\begin{eqnarray*}}
\newcommand{\ea}{\end{eqnarray*}}
\newcommand{\dagga}{{\phantom{\dagger}}}
\newcommand{\bR}{\mathbf{R}}

\newcommand{\bk}{\mathbf{k}}

\newcommand{\bRp}{\mathbf{R'}}

\newcommand{\Pj}[2]{|#1\rangle\langle #2|}

\newcommand{\fract}[2]{\frac{\displaystyle #1}{\displaystyle #2}}

\newcommand{\eqn}[1]{(\ref{#1})}

\begin{document}

\title{Surface dead layer for quasiparticles near a Mott transition}
\author{Giovanni Borghi}
\affiliation{International School for Advanced Studies (SISSA), and CRS Democritos, CNR-INFM,
Via Beirut 2-4, I-34014 Trieste, Italy} 
\author{Michele Fabrizio}
\affiliation{International School for Advanced Studies (SISSA), and CRS Democritos, CNR-INFM,
Via Beirut 2-4, I-34014 Trieste, Italy} 
\affiliation{The Abdus Salam International Centre for Theoretical Physics 
(ICTP), P.O.Box 586, I-34014 Trieste, Italy}
\author{Erio Tosatti}
\affiliation{International School for Advanced Studies (SISSA), and CRS Democritos, CNR-INFM,
Via Beirut 2-4, I-34014 Trieste, Italy} 
\affiliation{The Abdus Salam International Centre for Theoretical Physics 
(ICTP), P.O.Box 586, I-34014 Trieste, Italy}
\affiliation{Laboratoire de Physique des Solides, CNRS-UMR 8502, Universit\'e Paris-Sud, F-91405 Orsay, France}
\date{\today}
\begin{abstract}

Electron quasiparticles are progressively weakened by correlations upon approaching 
a continuos Mott metal insulator transition in a bulk solid. We show that corresponding 
to the bulk weakening, a dead layer forms below the surface of the solid, where 
quasiparticles are exponentially suppressed. The surface dead layer depth is a bulk
property, and diverges when the Mott transition is approached. We describe this phenomenon
in a Hubbard model within a self-consistent Gutzwiller approximation. Photoemission
data of Rodolakis {\sl et al.} in V$_2$O$_3$ appear to be in accord with this physical picture.

\end{abstract}

\maketitle

The Mott transition\cite{Mott} where a lattice of atoms or molecules abandons the 
metallic state and turns insulating due to
electron-electron repulsion, has a very 
intuitive physical explanation. Electron motion in the lattice is 
caused by kinetic energy, and favored by electron-ion energy because the same electron can 
feel in this way the attraction of more than one nucleus. It is opposed by Coulomb 
repulsion, higher for itinerant electrons due to the higher chance of collision
during motion. When the first two terms (which form the band
energy) prevail, the system is a band metal; otherwise 
the electrons localize, and we have an insulator. Despite that conceptual simplicity, 
properties of Mott insulators and especially of the strongly correlated metallic state 
close to a Mott transition remain quite difficult to capture both theoretically 
and experimentally. Theoretically, the reason is that the Mott transition is a collective 
phenomenon, which escapes single-particle or mean field theories such as Hartree-Fock 
or DFT-LDA approximations. Experimentally, complications 
such as magnetism, lattice distortions, etc., often conspire to mask the nature of metal 
insulator transitions.
 
Fresh progress on this problem has come in the last two decades with dynamical mean 
field theory (DMFT)\cite{DMFT}, which in the standard Hubbard model showed that, 
as the electron-electron repulsion parameter $U$ increases, the initial band-metal 
evolves first to a strongly correlated metal well before the Mott transition. 
In the strongly correlated metal the electron spectral function undergoes a 
profound change exhibiting well formed, localized Mott-Hubbard bands coexisting 
with delocalized, propagating quasiparticles -- the latter narrowly centered in 
energy near the Fermi level. Only successively do the quasiparticles
disappear as the Mott transition takes place when $U$ is increased to reach $U=U_c$. 
This intriguing prediction -- simultaneous metallic and insulating features, 
though on well separated energy scales --  
has stimulated a considerable experimental effort to reveal coexisting quasiparticles 
and Mott-Hubbard bands in strongly correlated metals\cite{Sekiyama,Maiti,Mo,Sekiyama-prl,Kamakura,Kim,Taguchi,Mo-prb,eguchi,Yano}. 
A large amount of work has been done on V$_2$O$_3$,
the prototype compound where a Mott transition was first discovered\cite{Remeika} and 
studied theoretically\cite{Rice,brinkman&rice}. 
At the metal-insulator transition of (V$_{1-x}$Cr$_x$)$_2$O$_3$, early photoemission 
experiments\cite{Post,Smith,Shin,Zimmermann} failed to reveal the sharp quasiparticle 
peak predicted by DMFT. The electronic spectrum was simply dominated by the lower 
Mott-Hubbard band with barely a hint of metallic weight at the Fermi energy.  
A similar puzzle was actually reported much earlier in $f$-electron materials\cite{Borje}, 
and soon ascribed to large surface effects in the presence of strong correlations\cite{Liu&Allen};     
the same conclusion reached by more recent photoemission 
experiments\cite{Sekiyama,Maiti,Kamakura,eguchi,panaccione,Sekiyama-prl,Yano}. 
In V$_2$O$_3$, using higher kinetic energy photo-electrons, whose escape depth is larger, 
a prominent quasiparticle peak coexisting with incoherent 
Mott-Hubbard bands was eventually observed \cite{vollardo,Mo,Mo-prb}.
Quasiparticle suppression in surface-sensitive probes was 
attributed\cite{vollardo} to surface-modified hamiltonian parameters,
the reduced atomic coordination pushing the surface 
closer to the Mott transition than the underlying bulk. 
Larger electronic correlations at the surface 
have been discussed by several authors through {\sl ad-hoc} formulations 
of DMFT\cite{Potthoff-2,Liebsch,ishida}. 
There is general agreement on intrinsically different quasiparticle properties near 
a surface, even if all hamiltonian parameters were to remain identically the same 
up to the outermost atomic layer\cite{Potthoff-2}.  

This conclusion, although not unexpected, raises a more fundamental question. 
A metal does not possess any intrinsic length-scale at long distances other than the 
Fermi wavelength. 
Thus an imperfection like a surface can only induce at large depth a power-law decaying 
disturbance such as that associated with Friedel's oscillations. 
Since one does not expect Luttinger's theorem to break down, 
even in a strongly correlated metal these oscillations should 
be controlled by the same Fermi wavelength as in the absence of interaction, 
irrespectively of the proximity of the Mott transition.   
However, a strongly correlated metal does possess an intrinsic energy scale, the parametric distance of the 
Hamiltonian from the Mott transition, where that distance could be associated with a 
length scale. The surface as a perturbation should alter the quasiparticle properties within a depth 
corresponding to that length, a bulk property increasing near the Mott transition, 
unlike the Fermi wavelength that remains constant. In this respect, it is not 
{\sl a priori} clear whether the recovery of bulk quasiparticles spectral 
properties with increasing depth should be strictly
power-law, compatible with the common view of a metal as an inherently critical state 
of matter, or whether it should be exponential, 
as one would expect by regarding the Mott transition  
as any other critical phenomena where power laws emerge only at criticality.   
We find here in the simple half-filled Hubbard model that the 
quasiparticle spectral weight below the surface is actually recovered exponentially
inside the bulk  with a length-scale that depends only on the bulk properties 
and diverges approaching the continuous Mott transition. 

To address the generic surface features of a 
a strongly correlated metal, we study the simplest Hamiltonian exhibiting a Mott transition, 
namely the Hubbard model at half-filling
 \be
H=-t\sum_{<\bR\bRp>\sigma}\,c^\dagger_{\bR\sigma}c^\dagga_{\bRp\sigma} + H.c. + \sum_\bR\,U_\bR\,n_{\bR\su}n_{\bR\giu},
\label{Ham}
\ee
where $<\bR\bRp>$ are nearest neighbor sites, $c^\dagger_{\bR\sigma}$ 
creates an electron at site $\bR$ with spin $\sigma$ and 
$n_{\bR\sigma}=c^\dagger_{\bR\sigma}c^\dagga_{\bR\sigma}$. Conventionally, the Mott transition 
of the half-filled Hubbard model is studied restricting to the paramagnetic sector of the Hilbert 
space\cite{Rice,brinkman&rice,DMFT} so as to avoid spurious effects due to magnetism. 
We assume a cubic lattice of spacing $a$ with periodic boundary conditions in $x$ and $y$ directions 
and open boundary conditions in the $z$ direction, in an $N$-layer slab geometry with two surfaces 
at $z=0$ and $z=N\,a$. The Hubbard electron-electron interaction parameter $U_\bR$ 
is $U$ everywhere except at the top surface layer($z=0$), where it takes 
a generally higher value $U_s>U$. In this way we can compare effects at the ideal lower surface 
($z=N\,a$), where $U_{Na}=U$, with the more correlated upper surface ($z=0$). 
DMFT\cite{DMFT} offers an ideal tool to attack this model in the 
paramagnetic sector, assuming a local self-energy that depends on the layer index 
$z$\cite{Potthoff-2,Liebsch,ishida}. 
However, a full DMFT calculation of this sort is numerically feasible only for a 
small number of layers, e.g. $N=20$ as in Ref.\cite{Rosch}, making the critical 
regime near the Mott transition hard to access.  As a useful approximate alternative, 
one can resort to the so-called linearized DMFT\cite{L-DMFT,Potthoff-2} to treat
moderately larger sizes. We decided to adopt a different 
method altogether, the Gutzwiller variational approximation\cite{mio}. 
Despite its limitations (static mean field character; inability to describe
the insulating phase) it is known to provide a good 
description of quasiparticle properties close to the Mott transition\cite{DMFT}
with very little size-limitations, and great simplicity and flexibility (it
may treat intersite interactions, any kind of lattice, etc.). 
We study \eqn{Ham} by means of a Gutzwiller type variational wavefuntion 
\be
|\Psi\rangle = \prod_\bR\,\mathcal{P}_\bR\, |\Psi_0\rangle,\label{GWF}
\ee
where $|\Psi_0\rangle$ is a paramagnetic Slater determinant. The operator $\mathcal{P}_\bR$ 
has the general expression
\be
\mathcal{P}_\bR = \sum_{n=0}^2\, \lambda_n(z)\,\Pj{n,\bR}{n,\bR},
\label{P_R}
\ee
where $\Pj{n,\bR}{n,\bR}$ is the projector at site $\bR=(x,y,z)$ onto configurations 
with $n$ electrons, and $\lambda_n(z)$ are layer-dependent variational parameters.  
We calculate average values 
on $|\Psi\rangle$ using the so-called Gutzwiller approximation\cite{Gutzwiller1,Gutzwiller2}, 
(for details see e.g. Ref.\cite{mio}, whose notations we use hereafter), and require that  
\be
\langle \Psi_0|\mathcal{P}^2_\bR|\Psi_0\rangle = 1,\;
\langle \Psi_0|\mathcal{P}^2_\bR\,n_{\bR\sigma}|\Psi_0\rangle = \langle \Psi_0|n_{\bR\sigma}|\Psi_0\rangle.\label{1}
\ee 
Because of particle-hole symmetry, $\langle \Psi_0|n_{\bR\sigma}|\Psi_0\rangle = 1/2$, from which it follows that 
Eq.~\eqn{1} is satisfied if 
$\lambda_2(z)=\lambda_0(z)$, $\lambda_1(z)^2 = 2 - \lambda_0(z)^2$.
The average value of \eqn{Ham} is then\cite{mio,Bunemann}
\bea
E&=&\fract{\langle \Psi|\,H\,|\Psi\rangle}{\langle \Psi|\Psi\rangle} = \sum_\bR\, \frac{U_\bR}{4}\,\lambda_0(z)^2
\label{E-var}\\
&& -t\sum_{<\bR\bRp>\sigma}\,R(z)\,R(z')\,
\langle \Psi_0 |c^\dagger_{\bR\sigma}c^\dagga_{\bRp\sigma} + H.c.|\Psi_0\rangle,\nonumber 
\eea
where $R(z) = \lambda_0(\bR)\sqrt{2-\lambda_0(\bR)^2}$
plays the role of a wavefunction renormalization factor. Its square is 
the actual quasiparticle weight, $Z(z)=R^2(z)$, 
since quasiparticle creation renormalizes into $R(z)\,c^\dagger_{\bR\sigma}$ in Fermi liquid theory.
One can invert this equation to express $\lambda_0(z)$ as function of $R(z)$, 
which become the actual variational parameters together with the Slater determinant $|\Psi_0\rangle$.   
In order to minimize $E$ in Eq.~\eqn{E-var} 
we assume that the Slater determinant $|\Psi_0\rangle$ is built with single-particle 
wavefunctions that, because of the slab geometry, have the general expression 
$
\phi_{\epsilon\bk_{||}}(\bR) = \sqrt{1/A}\,\mathrm{e}^{i\bk_{||}\cdot\bR}\, \phi_{\epsilon\bk_{||}}(z),
$
where $A$ is the number of sites per layer and $\bk_{||}$ the momentum in the $x$-$y$ plane.  
The stationary value of $E$ with respect to variation of $\phi_{\epsilon\bk_{||}}(z)$ 
and $R(z)$ corresponds to the coupled equations
\begin{widetext}
\bea
&&\epsilon\,\phi_{\epsilon\bk_{||}}(z) = R(z)^2\,\epsilon_{\bk_{||}}\, \phi_{\epsilon\bk_{||}}(z) 
- t\,R(z)\,\sum_{p=\pm}\, R(z+p\,a)\,\phi_{\epsilon\bk_{||}}(z+p\,a), \label{uno}\\
&&R(z) = \fract{4\sqrt{1-R(z)^2}}{U(z)A}\,\sum_{\epsilon\, \bk_{||}}^{occupied}\Bigg[-2\,R(z)\,
\epsilon_{\bk_{||}}\,\phi_{\epsilon\bk_{||}}(z)^2 
+t\,\phi_{\epsilon\bk_{||}}(z)\,\sum_{p=\pm}\, R(z+p\,a)\,\phi_{\epsilon\bk_{||}}(z+p\,a)\Bigg],\label{due}
\eea
\end{widetext}
where $\epsilon_{\bk_{||}}=-2t\left(\cos k_x a + \cos k_y a\right)$ and the sum  
in Eq.~\eqn{due} runs over all pairs of $\left(\epsilon,\bk_{||}\right)$ that are 
occupied in the Slater determinant $|\Psi_0\rangle$. The first equation has
the form of a Schr{\oe}dinger equation that the single-particle wavefunctions $\phi_{\epsilon\bk_{||}}(z)$ 
must satisfy, depending parametrically on $R(z)$.  The second equation has been 
intentionally cast in the form of a map $R_{j+1}(z)=F\left[R_j(z),R_j(z+a),R_j(z-a)\right]$ 
whose fixed point we have verified to coincide with the actual solution of 
\eqn{due} in the parameter region of interest. 
Eqs.~\eqn{uno} and \eqn{due} can be solved iteratively as follows. 
First solve the Schr{\oe}dinger equation at fixed $R_j(z)$; next find the 
new $R_{j+1}(z)$ using the old $R_j(z)$ and the newly determined 
wavefunctions $\phi_{\epsilon\bk_{||}}(z)$. With the new $R_{j+1}(z)$, 
repeat the above steps and iterate until convergence. Because of the large 
number of variational parameters, this iterative scheme 
is much more efficient than -- while fully equivalent to -- a direct minimization 
of $E$, Eq.~\eqn{E-var}.   

\begin{figure}
\includegraphics[width=8cm]{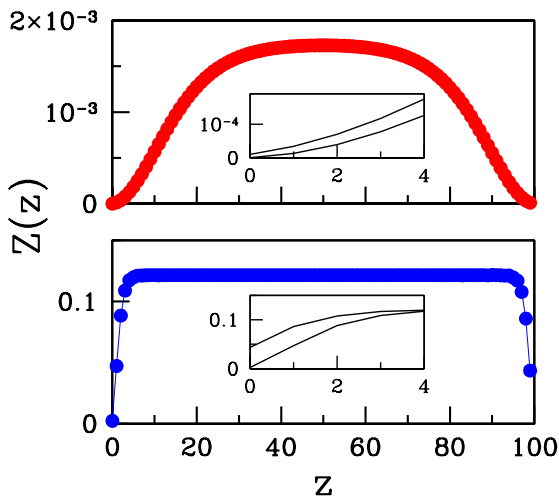}
\vspace{-1.2cm}
\caption{\label{fig1} (Color online) 
The quasiparticle weight $Z(z)=R^2(z)$ as function of the coordinate 
$z$ perpendicular to the surface (in units of the lattice spacing) for 
a 100-layer slab. The interaction parameter at $z=0$ is $U_s=20t$, while the bulk 
$U$ is $15.98t$ in the upper panel and $15t$ in the lower one (while $U_c$ =16). 
The insets show the behavior of $Z$ close to the two surfaces; the highest 
curve corresponding to the bulk-like surface, the other to $U_s = 20t$.}
\end{figure}    

\begin{figure}
\includegraphics[width=8.5cm]{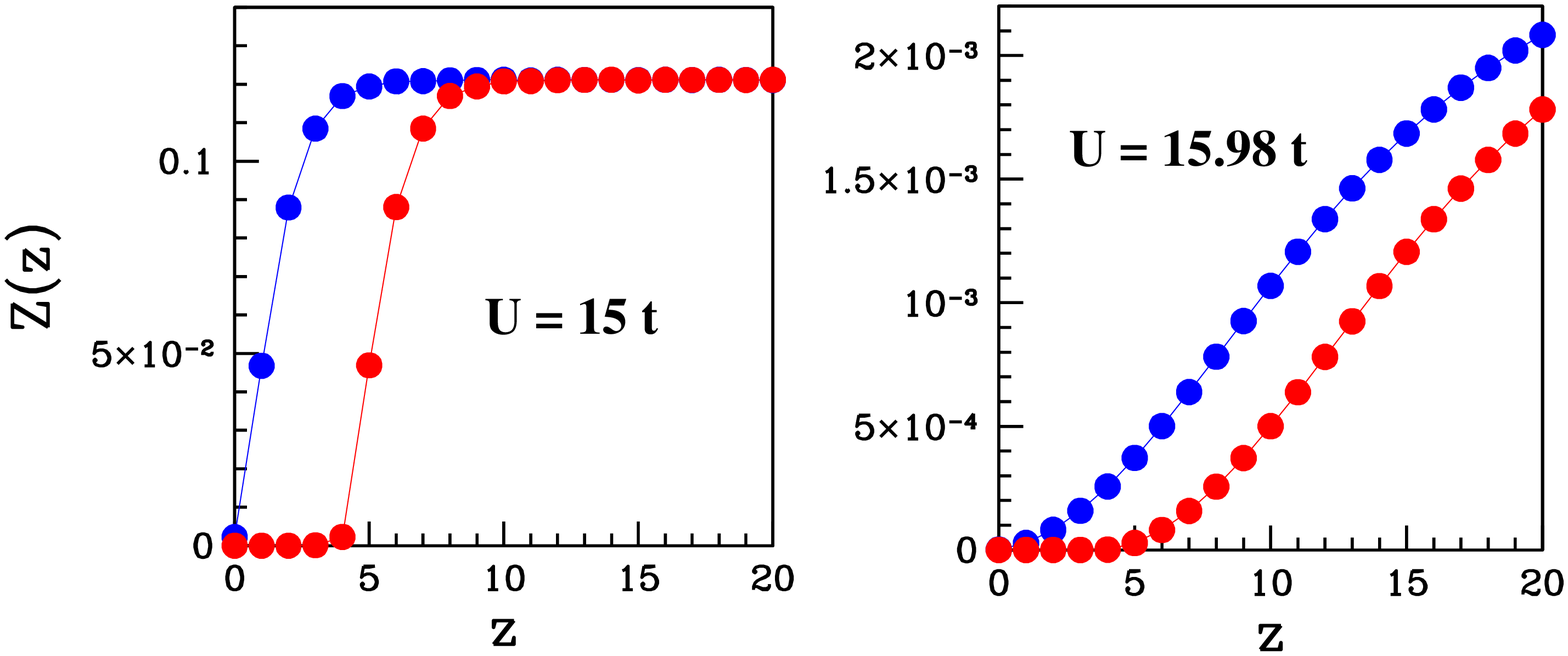}
\vspace{-1cm}
\caption{\label{fig2} Quasiparticle weight dependence on the distance $z$ from the surface 
for two different bulk $U$ values and for two cases: one where only the  
first layer has $U_s=20~t>U$ (upper curve in each panel), the other where 
five surface layers have $U_s=20~t$.} 
\end{figure}

\begin{figure}
\includegraphics[width=8cm]{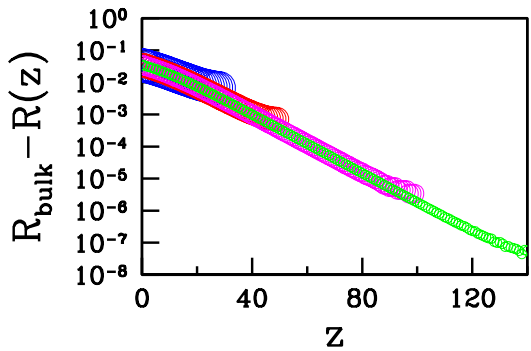}
\vspace{-3.2cm}
\caption{\label{fig3} Log scale plot of $R_{bulk}-R(z)$ versus $z$ for $U=15.99$, 
$U_s=20t$ and for different thicknesses of the slab $N=60,100,200,400$.}
\end{figure}

In Fig.~\ref{fig1} we plot $Z(z)=R^2(z)$, experimentally the total spectral 
weight carried by quasiparticles, calculated as function of $z$ (in units of the 
lattice spacing $a$) for $U_s=20t$, for two different bulk values 15$t$ and 15.98$t$ 
of $U$ below the critical Mott-transition value  $U_c=16t$.
Coming from the bulk, the quasiparticle weight $Z(z)$ decreases monotonically 
on approaching both surfaces, where it attains much smaller values than 
in bulk. As expected, the more correlated surface has a smaller quasiparticle
weight, $Z(0) < Z(N)$. Note however that so long as the slab interior (the ``bulk'')
remains metallic, the surface quasiparticle weight never vanishes
no matter how large $U_s$\cite{Potthoff-2}. Mathematically, this follows from Eq.~\eqn{due}, 
which is not satisfied by choosing $R(0)=0$ while $R(z>0)\not = 0$. Physically,
some metallic character can always tunnel from the interior to the surface,
so long as the bulk is metallic. 
The quasiparticle weight approaches the surface with upward curvature 
when $U$ is closest to $U_c$, upper panel in Fig.~\ref{fig1}, whereas
the behavior is linear well below $U_c$, as found earlier 
within linearized DMFT\cite{Potthoff-2}. We note that an upward curvature is in better 
accord with photoemission spectra of Rodolakis {\sl et al.} on  V$_2$O$_3$\cite{Marsi}. 
The curvature becomes more manifest if the number of surface layers where
$U_s>U$ is increased, as shown in Fig.~\ref{fig2}. 
Next, we analyse the dependence of $R(z)$ at large distance $ 1 << z << N/2$ 
below the surface.  As Fig.~\ref{fig3} shows, we find no trace of a power law,
and $R$ is best fit by an exponential 
$
R(z) = R_{bulk} + \bigg(R_{surf} - R_{bulk}\bigg)\,\mathrm{e}^{-z/\lambda}
$,
where $R_{bulk}$ is the bulk value (a function of $U$ only) and 
$R_{surf}<R_{bulk}$. $R_{surf}$ now depends on both $U$ and on $U_s$, and vanishes 
only when $R_{bulk}$ does at $U > U_c$. 
A detailed study by varying $U$ and $U_s$ shows that the surface ``dead layer''
thickness $\lambda$ depends only on bulk properties and diverges at the 
Mott transition as $\lambda \propto \left(U_c - U\right)^{-\nu}$. 
Numerically we find $\nu=0.53 \pm 0.3\simeq 0.5$, a typical mean field exponent\cite{Rosch}. 
The same conclusion can actually be drawn by analysing Eqs.~\eqn{uno} and \eqn{due} 
deep inside the bulk. We note that the precise behavior at the outermost 
surface layers would in a real system depend on details, such as 
lack of electron-hole symmetry and/or surface dipoles, not included in our 
model. 
However, we believe that the exponential behavior and its divergence
at a continuous Mott transition should be generic and universal, and thus 
independent of these and other details.
In conclusion, we have shown in a simple approximation the existence in 
the Hubbard model of strongly correlated metals of a ``dead layer''
below the crystal surface. Within this layer -- whose depth is a bulk property
and not a surface property of the metal -- the quasiparticle weight decays
exponentially on approaching the surface. The dead layer thickness $\lambda$
inversely depends on the distance in parameter space to the bulk continuous Mott transition, 
where it diverges critically. The physical significance of $\lambda$ is 
that of a correlation length of the bulk metallic state, where the quasiparticle 
weight acts as an order parameter, critically vanishing at a continuous Mott transition.
Like other features of the Hubbard model, this result should we believe carry 
over to real systems with an ideal Mott transition, not obscured by e.g., 
symmetry breaking phenomena like magnetic order, provided that the critical 
region is not preempted by a strong first order jump, like that in the 
$\alpha$-$\gamma$ transition of Ce. It could therefore apply 
to high temperature V$_2$O$_3$ near the paramagnetic metal-insulator weakly first order line,
notwithstanding complications including orbital degeneracy, Hund's rules, and 
coupling to the lattice (see e.g. Ref.\cite{poteryaev} and references therein).
We thus expect a surface dead layer in the metal phase of V$_2$O$_3$ , with
thickness increasing (although not diverging because of the first order transition) 
on approaching the Mott transition line. The associated paper by Rodolakis {\sl et al.} 
reports photoemission evidence which lends some support to this picture. It is also
interesting to note that an anomalously thick sub-surface dead layer 
has long been observed in mixed valent YbInCu$_4$\cite{Sato&Allen}, with a depth not smaller that 
60\AA\cite{moreschini}. 

We are indebted to Marino Marsi for sharing with us his data and for discussions which 
prompted this study. E.T. gratefully acknowledges the Laboratoire de Physique des Solides 
and the Universit\'e Paris-Sud for hospitality during the period where this work began. 
Work in SISSA was sponsored by PRIN-2006022847.


\end{document}